\documentclass[times]{aastex631}

\usepackage{amssymb}
\usepackage{amsbsy}
\usepackage{amsmath}
\usepackage{amsfonts}
\usepackage{mathrsfs}
\usepackage{color}
\usepackage{bm}

\usepackage{physics}
\usepackage{mathrsfs}
\let\tablenum\relax
\usepackage{siunitx}
\usepackage{lipsum}

\def\ihep{Key Laboratory for Particle Astrophysics, Institute of High Energy Physics, Chinese Academy of Sciences, 19B Yuquan Road, Beijing 100049, China}

\def\dgns{Dongguan Neutron Science Center, 1 Zhongziyuan Road, Dongguan 523808, China}

\def\UCAS{School of Astronomy and Physics, University of Chinese Academy of Sciences, 19A Yuquan Road, Beijing 100049, China}

\def\naoc{National Astronomical Observatories of China, Chinese Academy of Sciences, 20A Datun Road, Beijing 100020, China}

\def\USTC{Hefei National Research Center for Physical Sciences at the Microscale and School of Physical Sciences, University of Science and Technology of China, Hefei 230026, China}

\def\HNL{Hefei National Laboratory, University of Science and Technology of China, Hefei 230088, China}

\def\ynao{Yunnan Observatory, Chinese Academy of Sciences, Kunming 650011, Yunan, China}

\def\xmu{Department of Astronomy, Xiamen University, Xiamen, Fujian 361005, China}

\def\nbu{Institute of Fundamental Physics and Quantum Technology, Ningbo University, Ningbo, 315211, China}

\def\KeyLab{CAS Key Laboratory of Theoretical Physics, Institute of Theoretical Physics, Chinese Academy of Sciences (CAS), Beijing 100190, China}

\def\HIAS{School of Fundamental Physics and Mathematical Sciences, Hangzhou Institute for 
Advanced Study (HIAS), University of Chinese Academy of Sciences (UCAS), Hangzhou
310024, China}

\newcommand\cblue{\color{black}}

\begin{document}

\title{Geometrical Distances of Extragalactic Binaries through Spectroastrometry}

\correspondingauthor{Jian-Min Wang}
\email{wangjm@ihep.ac.cn}

\author[0000-0003-4042-7191]{Yu-Yang Songsheng}
\affiliation{\ihep}
\affiliation{\dgns}

\author[0000-0001-9449-9268]{Jian-Min Wang}
\affiliation{\ihep}
\affiliation{\UCAS}
\affiliation{\naoc}

\author[0000-0002-0354-2855]{Yuan Cao}
\affiliation{\USTC}
\affiliation{\HNL}

\author[0000-0001-5284-8001]{XueFei Chen}
\affiliation{\ynao}

\author[0000-0003-4829-6245]{JianPing Xiong}
\affiliation{\ynao}


\author[0000-0002-2419-6875]{Zhi-Xiang Zhang}
\affiliation{\xmu}

\author[0000-0002-3539-7103]{Rong-Gen Cai}
\affiliation{\nbu}
\affiliation{\KeyLab}
\affiliation{\HIAS}

\begin{abstract}
The growing ``Hubble tension'' has prompted the need for precise measurements of cosmological distances.
This paper {\cblue demonstrates a purely geometric} approach for determining the distance to extragalactic binaries through a joint analysis of spectroastrometry (SA), radial velocity (RV), and light curve (LC) observations. 
A parameterized model for the binary system is outlined, and simulated SA, RV, and LC data are computed to infer the probability distribution of model parameters based on the mock data. 
The impact of data quality and binary parameters on distance uncertainties is comprehensively analyzed, showcasing the method's potential for high-precision distance measurements. 
For a typical eclipsing binary in the Large Magellanic Cloud (LMC), the distance uncertainty is approximately $\SI{6}{\percent}$ under reasonable observational conditions. 
Within a specific range of data quality and input parameters, the distance measurement precision of individual binary star systems is generally better than $\SI{10}{\percent}$.
As a geometric method based on the simplest dynamics, it is independent of empirical calibration and {\cblue the systematics caused by model selections can be tested using nearby binaries with known distances}.
By measuring multiple binary star systems or monitoring one binary system repeatedly, geometric distance measurements of nearby galaxies can be achieved, providing valuable insights into the Hubble tension and advancing our understanding of the universe's structure and evolution. 
\end{abstract}

\keywords{}

\section{Introduction} \label{sec:intro}
The discrepancy in the Hubble constant obtained from measurements in the early and late universe \citep{planck2020b, riess2022}, known as the ``Hubble tension'', has become a significant concern due to advancements in measuring distances to local galaxies \citep{riess2021} and cosmic microwave background radiation \citep{planck2020a}.
This discrepancy has suggested that the standard $\Lambda$CDM model may require modifications, although it does not rule out the possibility of unknown systematic errors in the current measurements.
Over 100 schemes, including those involving dark energy, dark matter, modifications to gravity, inflation, cosmic phase transitions, and other new physics, have been put forward to address the Hubble tension \citep{divalentino2021}.
However, the testing of these models can only be achieved through observations, underscoring the importance of precise measurements of cosmological distances.

The distances of galaxies with redshifts less than 1 are typically measured through the cosmic distance ladder made by Cepheids and Type Ia supernovae.
The accuracy of the distance ladder is constrained by the uncertainty in calibrating the Cepheid period-luminosity relation \citep[e.g.,][]{freedman2010}.
Cepheids in the Milky Way can be calibrated using Gaia EDR3 parallaxes \citep{gaia2021}.
To investigate the relationship between the period-luminosity relation and factors such as metallicity abundance and other environmental influences \citep{freedman1990,sakai2004,fouque2007}, calibration with extragalactic Cepheids of known luminosities is necessary.
Hence, independent, reliable, and high-precision measurements of nearby galaxies' distances are essential.
Nevertheless, Gaia's capabilities do not extend to measuring the distances to the closest galaxies.
\citet{pietrzynski2013,pietrzynski2019} utilized the surface brightness-color relation (SBCR), calibrated using nearby red clump giant stars with known angular diameters \citep{gallenne2018}, to determine the angular sizes of red clumps in eclipsing binary systems from the Large Magellanic Cloud (LMC).
By integrating the physical sizes of stars obtained from eclipsing light curves (LC) and radial velocity (RV) curves, they attained a distance measurement to the LMC with $\SI{1}{\percent}$ precision.
In the case of more distant galaxies like M31, red clumps are too faint to be resolved. 
\citet{ribas} identified the early-type binary system V J00443799+4129236 in M31, consisting of O and B type stars, and determined its distance with an uncertainty of $\SI{5.7}{\percent}$ using similar methods. 
Lately, \citet{vilardell2010} measured the distance to the binary V J00443610+4129194 in M31 with an uncertainty of $\SI{4.4}{\percent}$.
Their measurements deviate by approximately $1-2\%$ from the distance obtained using Cepheids and remain consistent within the error bars. For blue stars with the color of $(V-K)_0\lesssim 2.0$, the SBCR has much larger scatters than red ones because of hot electron scattering \citep[e.g. their Figure 12 in][]{Taormina2019}. Moreover, factors like metallicity can result in subtle variations in the SBCRs observed in nearby stars compared to those in extragalactic regions \citep{salsi2022}.
This approach must also account for these potential systematic errors to improve the accuracy of distance measurements, particularly for hot stars.

In contrast, geometric methods offer unique advantages as they do not depend on specific empirical relationships requiring meticulous calibration.
Leveraging very long baseline interferometric observations of the positions, velocities, and accelerations of water masers in the active galaxy NGC 4258, \citet{reid2019} derived a geometric estimate of the angular-diameter distance to the galaxy with $\SI{1.5}{\percent}$ precision.
In the case of more distant active galaxies and quasars, a combined analysis of reverberation mapping and interferometry data can yield the physical and angular dimensions of the broad line regions or dusty torus surrounding their central massive black holes concurrently, thereby enabling the geometric determination of their distances \citep{honig2014,wang2020,gravity2021}.
Nevertheless, the majority of galaxies in the local group are inactive, necessitating the use of alternative standard rulers for geometric distance measurements.

Recently, \citet{gallenne2023} reported VLTI/GRAVITY observations of ten double-lined spectroscopic binaries in the Milky Way and determined their distances with an accuracy better than $\SI{0.1}{\percent}$ by combining RV measurements with interferometric observations. 
Extracting the angular separations between the stars in the binaries relies on measuring the interferometric visibilities, a task challenging for extragalactic binaries due to their angular separations of $\sim \SI{10}{\mu as}$, significantly beyond the resolution of optical interferometers in the foreseeable future.
{\cblue \citet{paczynski2001} proposed utilizing the Space Interferometry Mission (SIM) to achieve astrometric measurements with a precision of $\SI{1}{\mu as}$ on the light centroid of a visual binary in the Large Magellanic Cloud (LMC) for the determination of its geometric distance. 
Although SIM was ultimately suspended, VLTI/GRAVITY has successfully achieved a positioning accuracy of $\SI{10}{\mu as}$ through spectroastrometry\citep{gravity2017}. 
This approach involves measuring variations in the interferometric phases across emission or absorption lines, thereby revitalizing the prospects for purely geometric distance measurements of extragalactic binaries.}

In this paper, we {\cblue 
demonstrate a purely geometric approach} for determining the distance to extragalactic binaries through the integration of SA, RV, and LC observations.
In Section 2, we outline a parameterized model for the binary system, explaining the methodology for computing simulated SA, RV, and LC data using this model. Additionally, we discuss how to infer the probability distribution of model parameters based on observational data.
Section 3 showcases simulated observational data generated using the fiducial model parameters, along with the corresponding parameter reconstruction outcomes. 
We delve into a comprehensive analysis of the impact of data quality and binary parameters on distance uncertainties.
Section 4 is dedicated to discussing the challenges and considerations surrounding distance measurements.
Finally, we present our key findings and insights in the concluding section.

\section{Methods} \label{sec:methods}
\subsection{Spectroastrometry}
The SA measures the variations of angular displacements of photocentres along the projected baseline with wavelengths, which can be derived from the surface brightness distributions of the object \citep{lachaume2003, rakshit2015}.
For a binary star system, the surface brightness can be modeled as 
\begin{equation}
    \mathcal{O}(\vb*{\alpha},\lambda) = \sum_{i=1,2} \left[\mathcal{O}_{{\rm c},i}(\vb*{\alpha} - \vb*{\alpha}_i) - \mathcal{O}_{{\rm \ell},i}(\vb*{\alpha} - \vb*{\alpha}_i,\lambda)\right],
\end{equation}
where $\vb*{\alpha}$ is the angular coordinates on the plane tangent to the celestial sphere, $\lambda$ is the wavelength, $\mathcal{O}_{{\rm c}, i}$ and $\mathcal{O}_{{\rm \ell}, i}$ are the surface brightness of the continuum and absorption line of the $i$-th star, respectively.
The angular displacement of the photocentre is therefore
\begin{equation}
    \vb*{\epsilon}(\lambda) = \frac{\int \vb*{\alpha} \mathcal{O}(\vb*{\alpha},\lambda) \dd[2]\vb*{\alpha}}{\int \mathcal{O}(\vb*{\alpha},\lambda) \dd[2]\vb*{\alpha}} = \frac{\sum_{i=1,2} \left[ F_{{\rm c},i}\vb*{\epsilon}_{{\rm c},i} - F_{{\rm \ell},i}(\lambda)\vb*{\epsilon}_{{\rm \ell},i}(\lambda)\right]}{\sum_{i=1,2} \left[ F_{{\rm c},i} - F_{{\rm \ell},i}(\lambda)\right]}.
\end{equation}
Here, $F_{{\rm c}, i}$ and $F_{{\rm \ell}, i}$ are the flux of the continuum and absorption line of the $i$-th star:
\begin{equation}
    F_{{\rm c},i} = \int \mathcal{O}_{{\rm c},i}(\vb*{\alpha} - \vb*{\alpha}_i) \dd[2]\vb*{\alpha} \qc 
    F_{{\rm \ell},i}(\lambda) = \int \mathcal{O}_{{\rm \ell},i}(\vb*{\alpha} - \vb*{\alpha}_i,\lambda) \dd[2]\vb*{\alpha};
\end{equation}
and $\vb*{\epsilon}_{{\rm c}, i}$ and $\vb*{\epsilon}_{{\rm \ell}, i}$ are the corresponding angular displacement of the photocentre of the continuum and absorption line:
\begin{equation}
    \vb*{\epsilon}_{{\rm c},i} = \int \vb*{\alpha} \mathcal{O}_{{\rm c},i}(\vb*{\alpha} - \vb*{\alpha}_i) \dd[2]\vb*{\alpha} \qc
    \vb*{\epsilon}_{{\rm \ell},i}(\lambda) = \int \vb*{\alpha} \mathcal{O}_{{\rm \ell},i}(\vb*{\alpha} - \vb*{\alpha}_i,\lambda) \dd[2]\vb*{\alpha}.
\end{equation}
For simplicity, the SA observation is conducted when there is no eclipse, and so $\vb*{\epsilon}_{{\rm c}, i} = \vb*{\alpha}_i$.
Define the continuum flux ratio between the primary and secondary star as $\ell \equiv F_{{\rm c},1} / F_{{\rm c},2}$ and the ratio between the line flux and total flux as
\begin{equation}
    f_{{\rm \ell},i}(\lambda) \equiv \frac{F_{{\rm \ell},i}(\lambda)}{\sum_{i=1,2} \left[ F_{{\rm c},i} - F_{{\rm \ell},i}(\lambda)\right]},
\end{equation}
we have
\begin{equation}
    \vb*{\epsilon}(\lambda) = (1 + f_{{\rm \ell},1} + f_{{\rm \ell},2})\vb*{\epsilon}_{{\rm c}} - f_{{\rm \ell},1}\vb*{\epsilon}_{{\rm \ell},1} - f_{{\rm \ell},2}\vb*{\epsilon}_{{\rm \ell},2},
\end{equation}
where
\begin{equation}\label{eq:continuum_photon_center}
    \vb*{\epsilon}_{{\rm c}} = \frac{\ell \vb*{\alpha}_1 + \vb*{\alpha}_2}{\ell + 1},
\end{equation}
is the photocenter of the continuum flux.
At the reference wavelength $\lambda_{\rm r}$ where there is no absorption line, the displacement of photoncenter $\vb*{\epsilon}(\lambda_{\rm r})$ is only determined by that of the continuum photons:
\begin{equation}
    \vb*{\epsilon}(\lambda_{\rm r}) = \vb*{\epsilon}_{{\rm c}}.
\end{equation}
The differential phase curve, which measures the phase difference between the observed wavelength $\lambda$ and the reference wavelength $\lambda_{\rm r}$, is therefore
\begin{equation}\label{eq:diffphase}
    \Delta\phi(\lambda) = -\frac{2\pi}{\lambda} \vb*{B} \vdot \left[ \vb*{\epsilon}(\lambda) -  \vb*{\epsilon}(\lambda_{\rm r}) \right] 
    = -\frac{2\pi}{\lambda} \vb*{B} \vdot \left\{ 
    f_{{\rm \ell},1}\left[\vb*{\epsilon}_{{\rm c}} - \vb*{\epsilon}_{{\rm \ell},1}(\lambda)\right] 
    +  f_{{\rm \ell},2}\left[\vb*{\epsilon}_{{\rm c}} - \vb*{\epsilon}_{{\rm \ell},2}(\lambda)\right]\right\}.
\end{equation}
As shown by Eq. \ref{eq:diffphase}, the differential phase $\Delta\phi(\lambda)$ depends on the angular displacement of the photocenter of the absorption line $\vb*{\epsilon}_{{\rm \ell},i}(\lambda)$ relative to the photocenter of the continuum $\vb*{\epsilon}_{{\rm c}}$, weighted by the relative intensity of the line $f_{{\rm \ell},i}$. 
The continuum photocenter is the mean position of the two stars weighted by their luminosity.
If the luminosity ratio $\ell$ between the primary and secondary stars is equal to their mass ratio $q$, the continuum photocenter $\vb*{\epsilon}_{{\rm c}}$ will coincide with the mass center of the binary and keep stationary.
Otherwise, it will orbit around the mass center.
The profiles $\vb*{\epsilon}_{{\rm \ell},1}(\lambda)$ and $\vb*{\epsilon}_{{\rm \ell},2}(\lambda)$ will be redshifted and blueshifted respectively or vice versa, and the direction of $\vb*{\epsilon}_{{\rm \ell},1}$ and $\vb*{\epsilon}_{{\rm \ell},2}$ are opposite.
As a result, the differential phase is a S-shaped curve, whose amplitude and width are mainly determined by the angular separation and orbital velocity of the binary.

\subsection{Binary Orbit}
For a star in an elliptical orbit with eccentricity $e$ and period $P$, the eccentric anomaly $E$ at time $t$ can be obtained by solving the following Kepler \citep{heihtz1978},
\begin{equation}
    2\pi\left(\frac{t-T_0}{P}\right) = E - e\sin E,
\end{equation}
where $T_0$ is the time passage through the periastron.

\begin{figure}
    \plotone{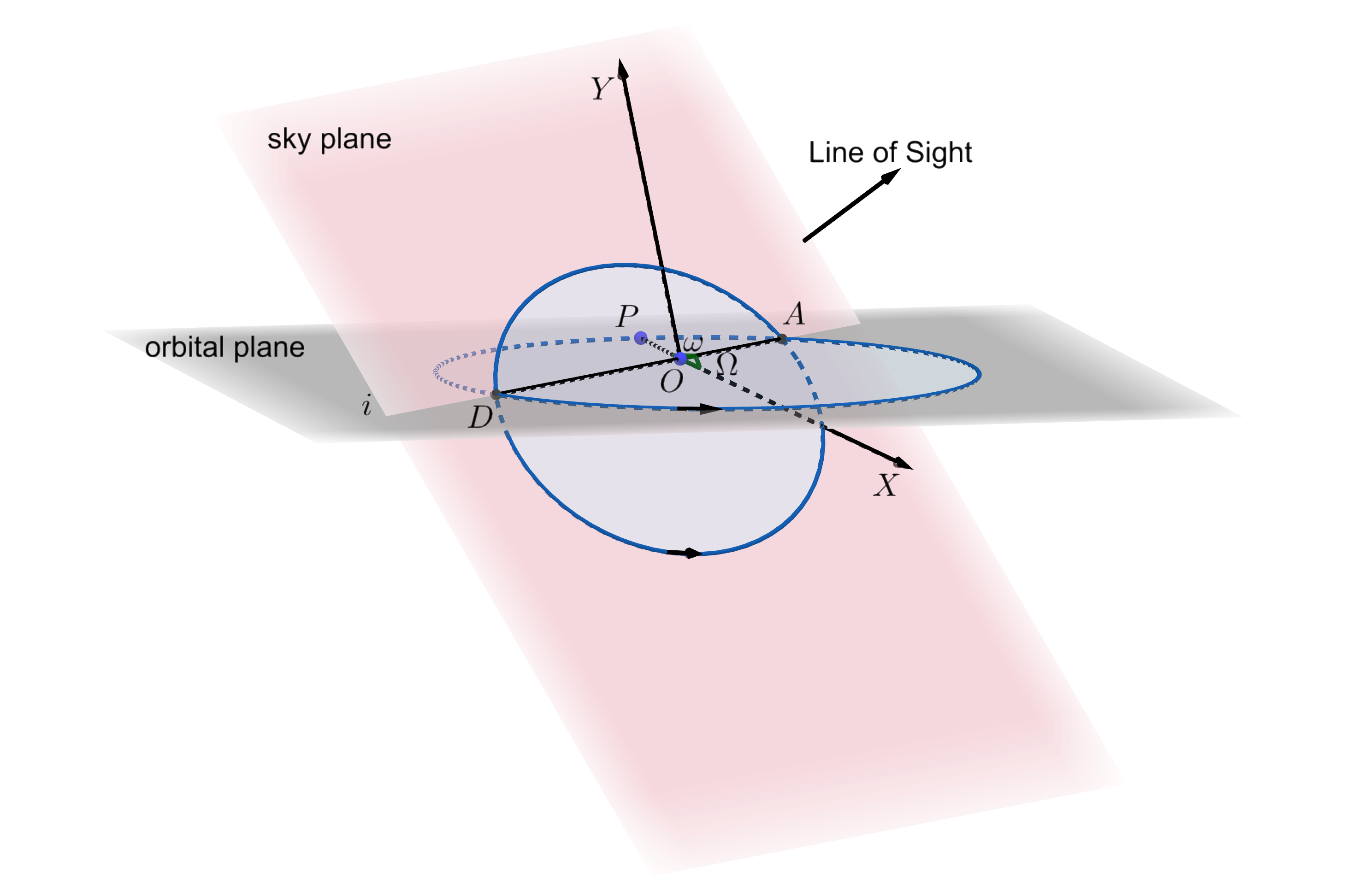}
    \caption{\footnotesize The orbit of the binary and its projection onto the celestial plane.
    The focal point $O$ of the elliptical orbit is chosen as the coordinate origin. The orbital plane and the celestial plane intersect at line $AD$ with an angle $i$, where $A$ and $D$ represent the ascending and descending nodes respectively. In the celestial plane, $OX$ points north, and $OY$ points east. $\Omega$ represents the azimuthal angle of ascending node $A$ relative to the north direction $OX$, while $\omega$ denotes the angle from $A$ to the periastron $P$.}
    \label{fig:orbit}
\end{figure}

The orbit of the binary and its projection onto the celestial plane is illustrated by Fig. \ref{fig:orbit}.
To project the true orbit onto the celestial sphere, we define the following Thiele-Innes elements as
\begin{align}
    A = a(\cos\omega\cos\Omega - \sin\omega\sin\Omega\cos i) \qc B = a(\cos\omega\sin\Omega + \sin\omega\cos\Omega\cos i), \nonumber \\
    F = a(-\sin\omega\cos\Omega - \cos\omega\sin\Omega\cos i) \qc G = a(-\sin\omega\sin\Omega + \cos\omega\cos\Omega\cos i),    
\end{align}
where $a$ is the semimajor axis of the relative orbit of the secondary star relative to the primary star, $i$ is the orbital inclination, $\Omega$ is the position angle of the ascending node and $\omega$ is the argument of periastron.
So the relative angular position of the secondary to the primary $\vb*{\alpha}_0 = (\alpha_x, \alpha_y)$ is given by
\begin{equation}
    \alpha_x = \frac{A}{D}\left(\cos E - e\right) + \frac{F}{D}\sqrt{1-e^2}\sin E \qc
    \alpha_y = \frac{B}{D}(\cos E - e) + \frac{G}{D}\sqrt{1-e^2}\sin E,
\end{equation}
where $D$ is the distance of the binary, and $\alpha_x$ and $\alpha_x$ point to the direction of increasing declination and right accession respectively.
Putting the origin at the center of mass of the binary, the angular position of the two stars are
\begin{equation}
    \vb*{\alpha}_1 = -\frac{1}{1+q}\vb*{\alpha}_0 \qc \vb*{\alpha}_2 = \frac{q}{1+q}\vb*{\alpha}_0,
\end{equation}
where $q$ is the mass ratio between the primary and secondary.

To obtain the RVs of the two stars, we further define
\begin{equation}
    C = a\sin\omega \sin i \qc H = a \cos\omega \sin i.
\end{equation}
The relative RV of the secondary to the primary is given by
\begin{equation}
    V_0 = \frac{2\pi}{P\left(1-e\cos E\right)}\left(-C\sin E + H \sqrt{1-e^2}\cos E\right),
\end{equation}
where $v_0 > 0$ means the star moves away from the observer.
So the RV of the two stars are 
\begin{equation}
    V_1 = -\frac{1}{1+q}V_0 \qc V_2 = \frac{q}{1+q}V_0.
\end{equation}
These projected velocities measured by spectrograph are more sensitive to the mass ratio of the binary stars, in particular the radial velocity ratio.

\subsection{Star surface brightness}
For simplicity, we model the surface of the star as a uniform disk where there is no eclipse, i.e.,
\begin{equation}
    \mathcal{O}_{{\rm c},i}(\vb*{\beta}_i) =  \begin{cases}
        {F_{{\rm c},i}}/{\pi \theta_i^2} \qif |\vb*{\beta}_i| \le \theta_i \\
        0 \qif |\vb*{\beta}_i| > \theta_i,
    \end{cases}
\end{equation}
where $\theta_i$ is the angular radius of the $i$-th star and $\vb*{\beta}_i = \vb*{\alpha}- \vb*{\alpha}_i$ is the relative angular displacement to the disk center.

For the absorption line, we include the rotation of the star and the thermal broadening, i.e.,
\begin{equation}
    \mathcal{O}_{{\rm \ell},i}(\vb*{\beta}_i,\lambda) = \frac{F_{{\rm \ell},i}}{\pi \theta_i^2} \frac{1}{\sqrt{2\pi}\sigma_i} \exp{-\frac{[\lambda - \lambda_i(\vb*{\beta}_i)]^2}{2\sigma_i^2}},
\end{equation}
where
\begin{equation}
    \lambda_i(\vb*{\beta}_i) = \lambda_0\left[1 + \frac{V_i}{c} + \frac{v_{{\rm rot},i}|\vb*{\beta}_i|\sin(\phi_{{\rm rot},i} - \phi_{\alpha})}{c\theta_i} \right],
\end{equation}
$\sigma_i$ is the thermal broadening of the absorption line, $\lambda_0$ is the central wavelength, $c$ is the speed of light, $v_{{\rm rot}, i}$ is the projected rotational velocity, and $\phi_{{\rm rot}, i}$ and $\phi_{\alpha}$ are the position angle of the projected rotational axis and relative angular displacement $\vb*{\beta}_i$ respectively.

\subsection{Eclipse}
The eclipse happens when the sum of two stars' radii is larger than the distance between them, i.e. $|\vb*{\alpha}_0| < \theta_1 + \theta_2$.
When the secondary star is obscured, the flux from it will be 
\begin{equation}
    F_{{\rm c},2}^{\prime} = \iint_{|\vb*{\beta}_2 + \vb*{\alpha}_0| > \theta_1} \mathcal{O}_{{\rm c},2}(\vb*{\beta}_2) \dd[2]\vb*{\beta}_2.
\end{equation}
For a uniform disk, we can figure out that
\begin{equation}
    F_{{\rm c},2}^{\prime} = F_{{\rm c},2} \left(1 - \frac{\Theta_2 - \cos\Theta_2\sin\Theta_2}{\pi} - \frac{\theta_1^2}{\theta_2^2} \cdot \frac{\Theta_1 - \cos\Theta_1\sin\Theta_1}{\pi}  \right)
\end{equation}
when $|\vb*{\alpha}_0| > |\theta_1 - \theta_2|$, where
\begin{equation}
    \Theta_1 = \arccos{\frac{|\vb*{\alpha}_0|^2 + \theta_1^2 - \theta_2^2}{2|\vb*{\alpha}_0| \theta_1}} \qc
    \Theta_2 = \arccos{\frac{|\vb*{\alpha}_0|^2 + \theta_2^2 - \theta_1^2}{2|\vb*{\alpha}_0| \theta_2}}.
\end{equation}
When $|\vb*{\alpha}_0| \le |\theta_1 - \theta_2|$, we have $F_{{\rm c},2}^{\prime} = 0$ if $\theta_1 \ge \theta_2$ and $F_{{\rm c},2}^{\prime} = F_{{\rm c},2}(1 - {\theta_1^2}/{\theta_2^2})$ if $\theta_1 < \theta_2$.
When the primary star is obscured, the calculation of the flux can be carried out similarly.
If the star is described by quadratic or nonlinear limb darkening rather than a uniform disk, the exact analytic formulae for the eclipse can be found in \citet{mandel2002}.

\subsection{Sampling}

\begin{table}
\footnotesize
\centering
\caption{Parameters used in the binary star model \label{tab:parameters}}
\begin{tabular}{llllll}\hline\hline
	Parameter & Unit & Implication & Fiducial Value & Range & Prior \\ \hline
    Orbits & & & & & \\ 
    $D$ & kpc & distance & $50$ & $[1,10^3]$ & log-uniform \\
    $P$ & day & orbital period  & $300$ &  & fixed \\
    $T_0$ & day & time of periastron & $0$ & & fixed \\
    $a$ & au & semi-major axis  & $2$ & $[0.1, 10]$ & log-uniform \\
    $e$ &  & eccentricity  & $0.3$ & $[0, 0.95]$ & uniform \\
    $i$ & $^\circ$ & orbital inclination & $87.13$ & $[0,180]$ & $\cos i$ uniform \\
    $\Omega$ & $^\circ$ & position angle & $45$ & $[0,360]$ & uniform \\
    $\omega$ & $^\circ$ & argument of periastron & $30$ & $[0,360]$ & uniform \\
    $q$ & & mass ratio & $1.5$ & $[1,100]$ & log-uniform \\
    $\ell$ & & flux ratio & $3$ & $[0.1, 100]$ & log-uniform \\ \hline
    Stars & & & & & \\
    $r_{i}$ & $a$ & size & $0.1$, $0.08$ & $[10^{-3}, 1]$ & log-uniform \\
    $v_{{\rm rot},i}$ & km/s & projected rotational velocity & $10$, $5$ & $[0.01, 100]$ & log-uniform \\
    $\phi_{{\rm rot},i}$ & $^\circ$ & position angle of rotational axis & $120$, $60$ & $[0, 360]$ & uniform \\
    $\sigma_i$ & km/s & thermal broadening & $10$, $10$ & $[0.01, 100]$ & log-uniform \\
    ${\rm EW}_i$ & \AA & equivalent width & $1$, $1$ & $[0.1, 10]$ & log-uniform \\
    \hline
\end{tabular}
\tablecomments{\footnotesize
Fiducial values follow the sequence of the 1st-star and 2nd-star.} 
\vglue 0.5cm
\end{table}

The model parameters ($\vb*{\Theta}$) for calculating SA, RV, and LC are summarized in Table \ref{tab:parameters}.
Given SA, RV, and LC data ($\mathscr{D}$), we can reconstruct the posterior probability distribution of model parameters in the Bayesian framework:
\begin{equation}
    P(\vb*{\Theta} | \mathscr{D}) = \frac{P(\mathscr{D} | \vb*{\Theta}) P(\vb*{\Theta}) }{P(\mathscr{D})},
\end{equation}
where $P(\vb*{\Theta})$ is the prior distribution of model parameters, $P(\mathscr{D})$ is a normalization factor and $P(\mathscr{D} | \vb*{\Theta})$ is the likelihood for given data and parameters.
We assume the prior probabilities of model parameters are independent. 
Each parameter's prior range and distribution are presented in Table \ref{tab:parameters}.

Suppose all measurements are independent and errors are Gaussian distributed.
The likelihood of the data is given by
\begin{align}
    P(\mathscr{D} | \vb*{\Theta}) &= \prod_{i=1}^{N_{\rm t,sa}} \prod_{j=1}^{N_{\lambda}} \prod_{k=1}^{N_{\rm b}} \frac{1}{\sqrt{2\pi}\sigma_{\phi,ijk}} \exp\left[ -\frac{(\phi_{ijk} - \phi_{ijk}^{\rm mod})^2}{2\sigma_{\phi,ijk}} \right] 
    \times \prod_{i=1}^{N_{\rm t,sa}} \prod_{j=1}^{N_{\lambda}}  \frac{1}{\sqrt{2\pi}\sigma_{{\rm \ell},ij}} \exp\left[ -\frac{(F_{{\rm \ell}, ij} - F_{{\rm \ell},ij}^{\rm mod})^2}{2\sigma_{{\rm \ell},ij}} \right] \nonumber \\
    &\times \prod_{i=1}^{N_{\rm t,rv}} \prod_{j=1}^{2} \frac{1}{\sqrt{2\pi}\sigma_{{\rm v},ij}} \exp\left[ -\frac{(V_{ij} - V_{ij}^{\rm mod})^2}{2\sigma_{{\rm v},ij}} \right] 
    \times \prod_{i=1}^{N_{\rm t,lc}} \frac{1}{\sqrt{2\pi}\sigma_{{\rm c},i}} \exp\left[ -\frac{(F_{{\rm c}, i} - F_{{\rm c},i}^{\rm mod})^2}{2\sigma_{{\rm c},i}} \right],
\end{align}
where $\phi_{ijk}$, $\phi_{ijk}^{\rm mod}$ and $\sigma_{\phi,ijk}$ are measured value, predicted value and measurement uncertainty of differential phase at the $i$-th time slot, $j$-th wavelength channel, and $k$-th baseline; 
$F_{\ell,ij}$, $F_{\ell, ij}^{\rm mod}$ and $\sigma_{\ell,ij}$ are measured value, predicted value and measurement uncertainty of the absorption line at the $i$-th time slot and $j$-th wavelength channel;
$V_{ij}$, $V_{ ij}^{\rm mod}$ and $\sigma_{{\rm v},ij}$ are measured value, predicted value and measurement uncertainty of the radial velocity of the $j$-th star at the $i$-th time slot;
$F_{{\rm c},i}$, $F_{{\rm c},i}^{\rm mod}$ and $\sigma_{{\rm c},i}$ are measured value, predicted value and measurement uncertainty of the ecplipsing light curve at the $i$-th time slot.

After establishing the formulations for the prior distribution of model parameters and the likelihood function, we can utilize appropriate Monte Carlo techniques \citep{sharma2017} to sample the posterior probability distribution of the parameters.
Specifically, we employ the Diffusive Nested Sampling algorithm \citep{brewer2011} for this purpose. 
Noteworthy for its ability to navigate intricate scenarios including high-dimensional parameter spaces and multimodal distributions, this algorithm also facilitates the computation of model evidence essential for model contrasts and comparisons.

\begin{figure}[ht!]
    \plotone{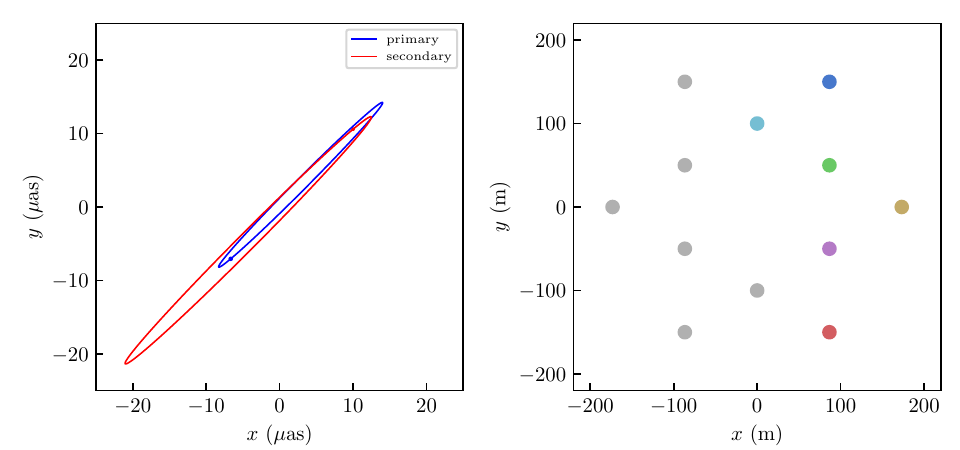}
    \caption{\footnotesize The left panel shows the projected orbits of both stars in the binary system. 
    The blue and red circles mark the sizes and the initial positions of the primary and secondary stars respectively.
    The right panel shows the six projected baselines we used for the simulation in the UV plane.
    We keep the projected baselines fixed for simplicity. 
    In practice, they will vary as the earth rotates.
    \label{fig:orbit_baseline}}
\end{figure}

\section{Results} \label{sec:results}
\subsection{Fiducial case}
In order to study the impact of data quality and model parameters on distance measurement, we must construct a fiducial case as a benchmark for comparison.
Table \ref{tab:parameters} lists model parameters for simulating the fiducial data set.
The resulting projected orbits of both stars are shown in the left panel of Fig. \ref{fig:orbit_baseline}.

The SA observation is taken at a cadence of $\SI{41}{days}$ lasting for $\SI{1}{year}$.
Eight observations were generated after removing the observations at the time of the two eclipsing events.
For simplicity, we keep the six projected baselines fixed in the UV plane during the observation, as shown in the right panel of Fig. \ref{fig:orbit_baseline}.
The lengths of baselines ranges from $\SI{100}{m}$ to $\SI{200}{m}$.
The spectral line for SA observation is Br$\gamma$ centered at $\SI{2.166}{\mu m}$ in the rest frame.
The simulated line profiles and differential phase curves are all convolved by a Gaussian with ${\rm FWHM} = \SI{30}{km \per s}$ (spectral resolution $\mathcal{R} = 10000$) to account for instrument broadening.
The bin size of the phase curve and spectra is half of the standard deviation of the Gaussian broadening function, i.e., $\SI{0.46}{\angstrom}$ in this case.
The uncertainties of phase and flux measurements in each wavelength channel are $\ang{0.1}$ and $\SI{1}{\percent}$ respectively.
The RV observation is taken at a cadence of $\SI{15}{days}$ lasting for $\SI{1}{year}$.
Two observations during the eclipse are also removed.
The uncertainty of the velocity measurement is $\SI{2}{km \per s}$ for each star.
The LC observation is taken at a cadence of $\SI{0.5}{day}$ lasting for $\SI{1}{year}$, and the uncertainty of flux measurement is $\SI{0.2}{\percent}$.
In practice, we can always increase the cadence and precision of the observation by period folding.

\begin{figure}
    \plotone{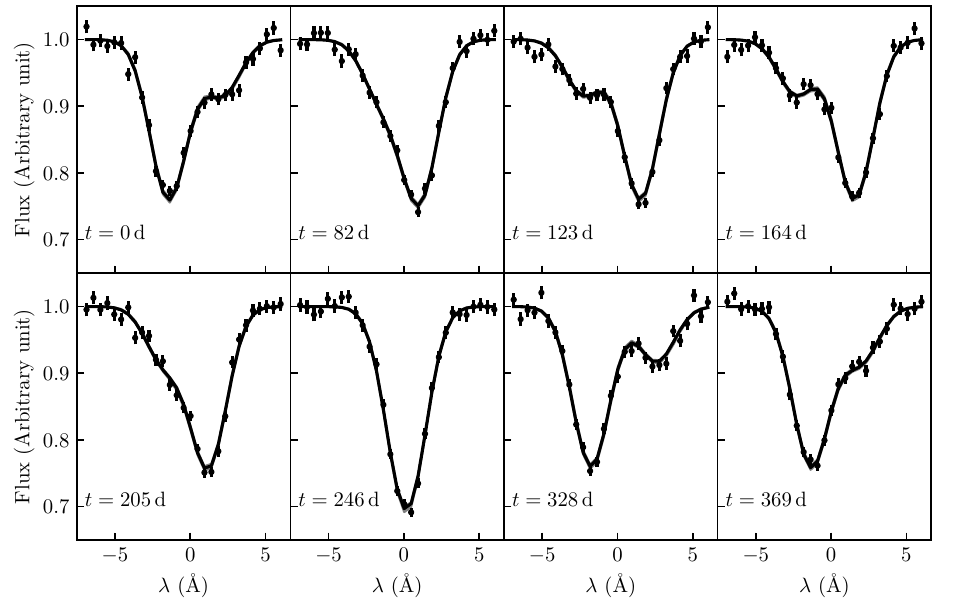}
    \caption{\footnotesize Profiles of the absorption line used for SA observation at different orbital phases. Black dots are measured values of line flux with $1\sigma$ error bars and black lines are best-fitted profiles.
    The continuum flux is normalized to $1$ and the center of the line is moved to $\SI{0}{\angstrom}$ for clarity.
    \label{fig:spec}}
\end{figure}

\begin{figure}[ht!]
    \includegraphics[width=0.97\textwidth]{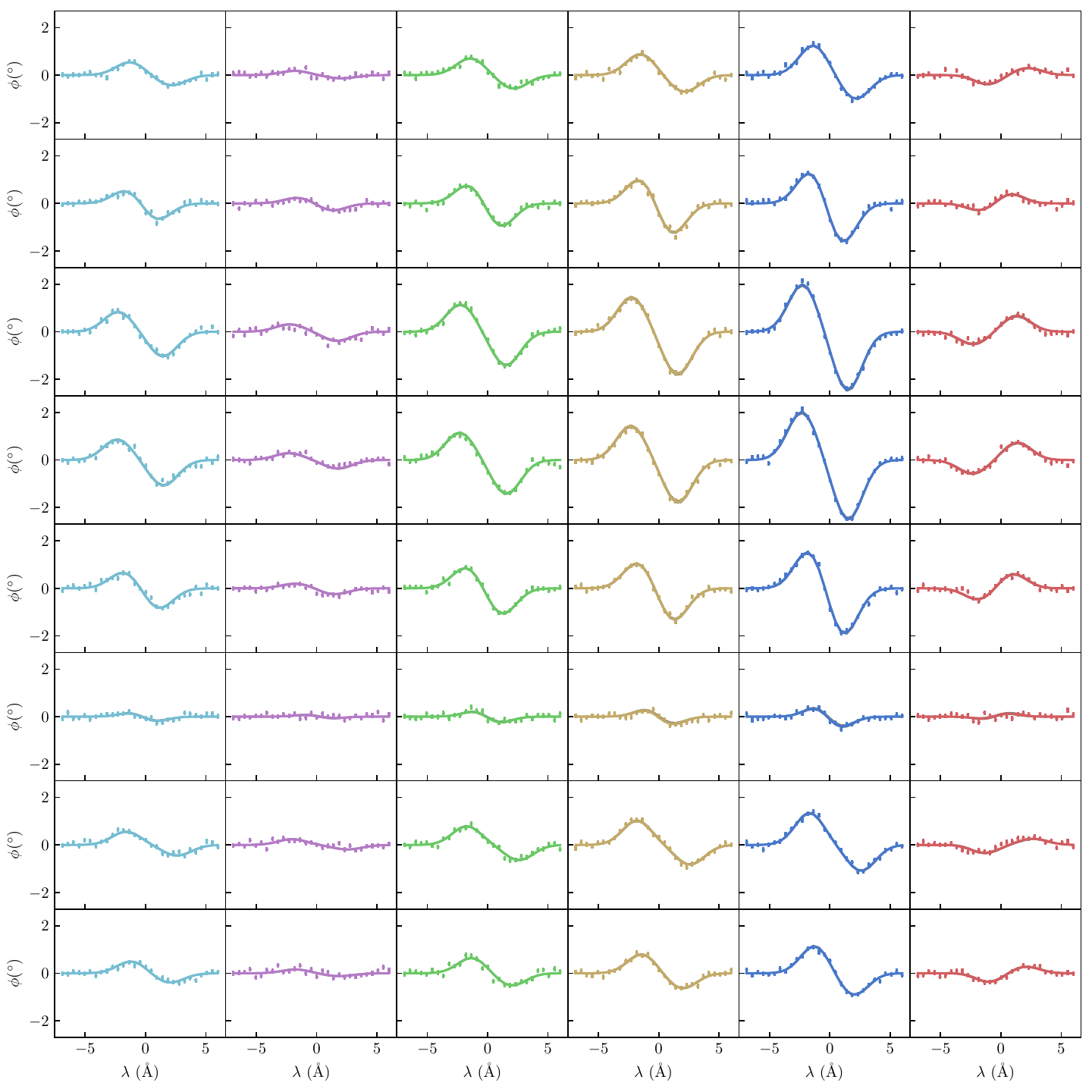}
    \caption{\footnotesize The differential phase measured by each baseline at different orbital phases and their best-fitted curves.
    The data points' colors correspond to the baselines' colors in the right panel of Fig. \ref{fig:orbit_baseline}.
    \label{fig:phase}}
\end{figure}

\begin{figure}[ht!]
    \plotone{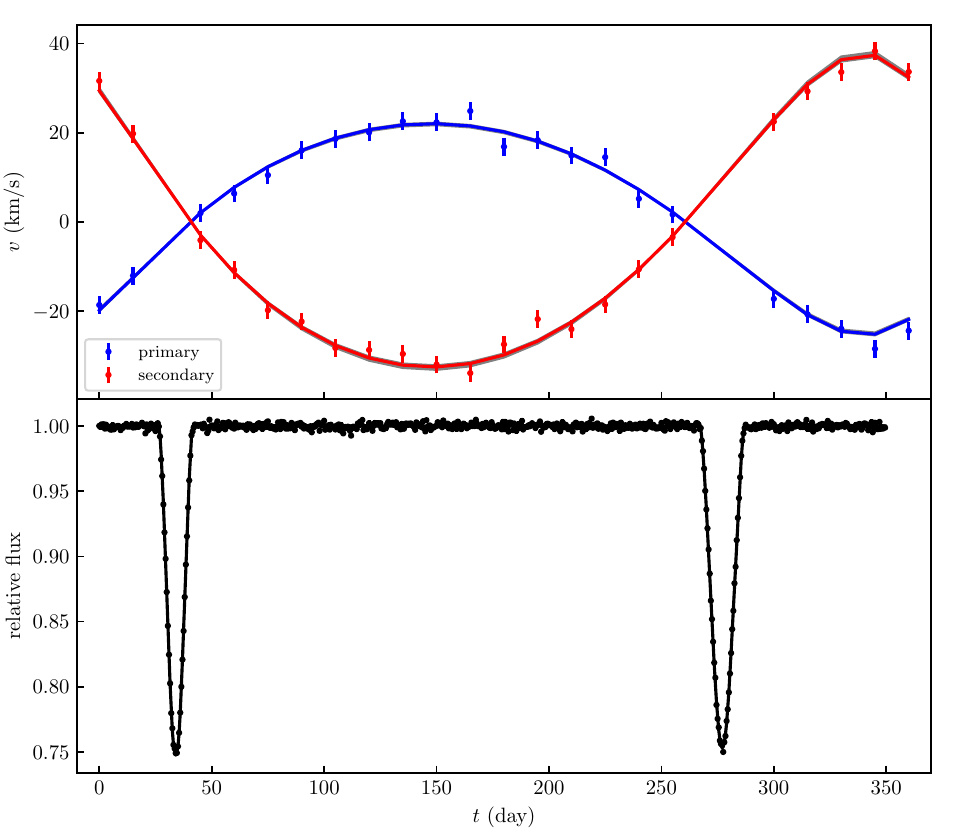}
    \caption{\footnotesize The upper panel shows the measured RV of each star and their best-fitted curves.
    The lower panel shows the measured LC and its best-fitted curve. 
    Flux drops when the eclipse happens.
    \label{fig:rvlc}}
\end{figure}

\begin{figure}[ht!]
    \includegraphics[width=0.99\textwidth]{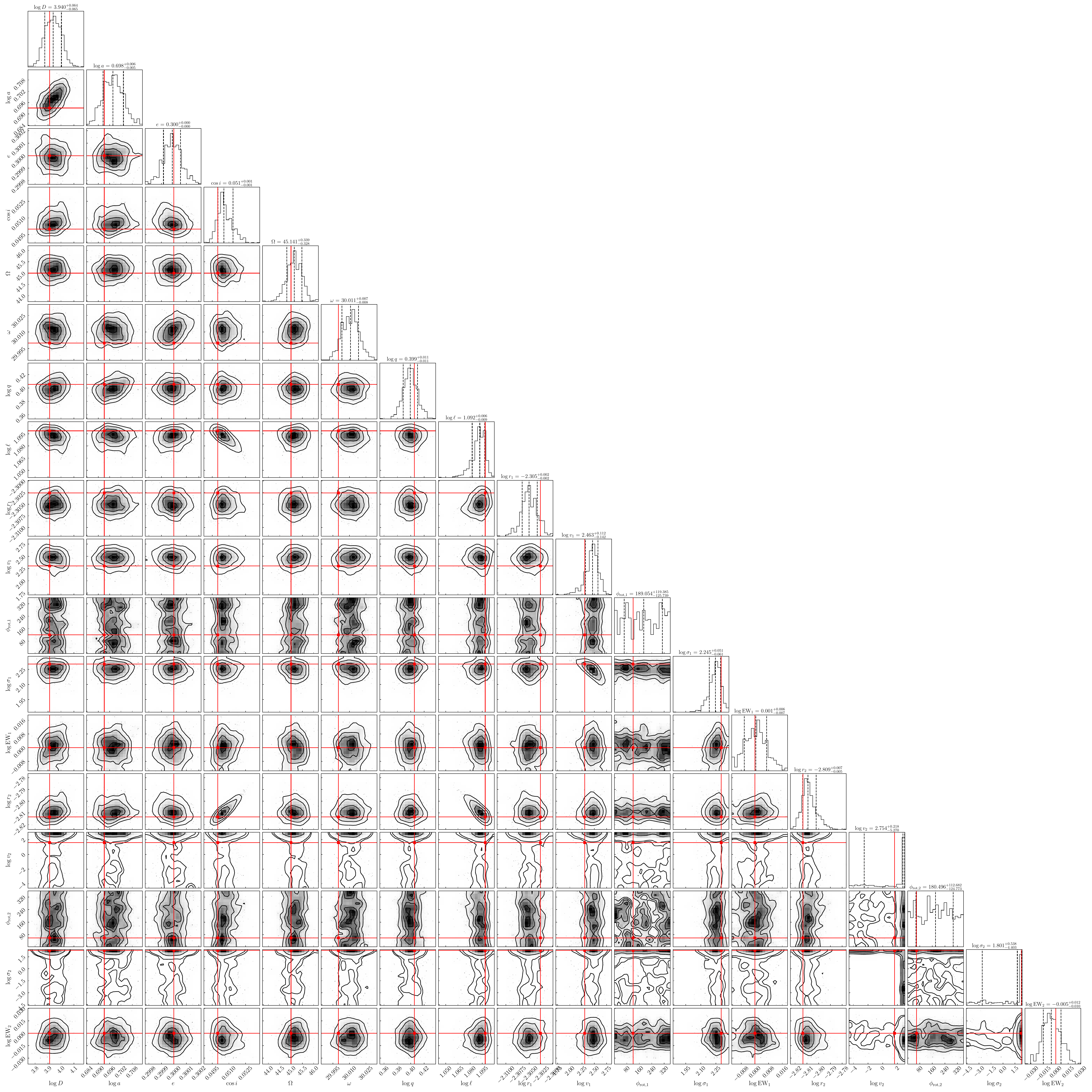}
    \caption{\footnotesize Posterior distributions of model parameters obtained by fitting the binary star model to the mock data. 
    The median values with error bars at $1\sigma$ level of all parameters are given on the tops of panels. Contours in two-dimensional distribution are at $1\sigma$, $1.5\sigma$, and $2\sigma$, respectively.
    The red lines represent input values.
    \label{fig:dis}}
\end{figure}

Then we will apply the diffusive nested sampling to fit the binary model to the mock data and obtain the probability distribution of model parameters.
In practice, the orbital period and time passage through periastron can always be determined precisely from long-term monitoring of the binary, and so will be fixed here for simplicity.
The simulated and reconstructed line profile, differential phase curve, RV, and LC are shown in Fig. \ref{fig:spec}, \ref{fig:phase}, and \ref{fig:rvlc} respectively.
As we can see, all mock data have been fitted very well within the margin of error.
Posterior distributions of model parameters are shown in Fig. \ref{fig:dis}.
Firstly, the relative uncertainty of the distance is about $\SI{6.5}{\percent}$ and the input value is within the $1\sigma$ range of the distribution.
The degeneracy between distance and other parameters is weak and so the uncertainty can be further reduced by improving the quality of the data without introducing significant bias.
Other orbital parameters are determined within $\SI{1}{\percent}$ for dimensional parameters and $\ang{1}$ for angles.
For star parameters, the relative sizes, thermal velocities, and equivalent widths are well determined, but velocities and directions of rotations are poorly constrained due to their degeneracies with thermal broadening under current spectral resolution.

\subsection{Dependence on data quality}
\begin{figure}[ht!]
    \includegraphics[width=0.95\textwidth]{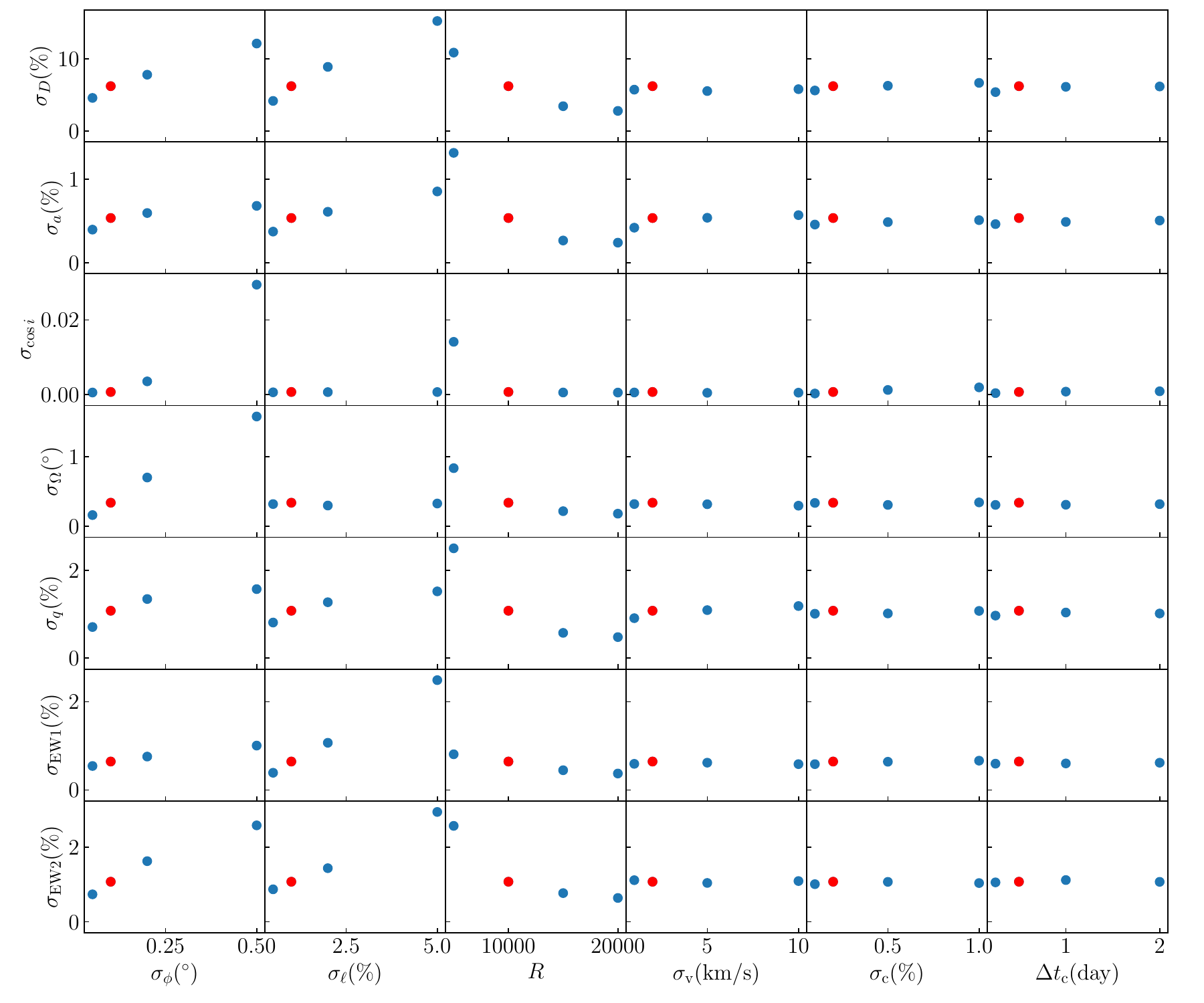}
    \caption{\footnotesize Dependence of measurement uncertainties on data quality for some parameters. 
    In each row, we depict the relationship between the uncertainties of a parameter and the quality of various datasets. 
    The red point denotes the uncertainties of the parameters derived from the fiducial mock data. 
    The uncertainties of distances decrease when we reduce the error bars of differential phases and line fluxes or enhance the spectral resolution, ranging from $\SI{2}{\percent}$ to $\SI{10}{\percent}$.
    They exhibit little sensitivity to the error bars of radial velocities and continuum fluxes, as well as the cadence of the eclipsing light curves.
    }
    \label{fig:std1}
\end{figure}

After analyzing the fiducial case, we will vary the quality of different data to study its impact on the uncertainties of model parameters, especially the uncertainty of binary distance.
The simulation results are summarized in Fig. \ref{fig:std1}.
As shown by the first row of Fig. \ref{fig:std1}, the relative uncertainty of the distance ($\sigma_{D}$) is approximately proportional to the error of differential phases ($\sigma_{\phi}$) and absorption line flux ($\sigma_{\rm \ell}$), inversely proportional to spectral resolution of SA observation ($\mathcal{R}$), but less dependent on the error of RV ($\sigma_{\rm v}$), LC ($\sigma_{\rm c}$) and the cadence of LC ($\Delta t_{\rm c}$).
As $\sigma_{\phi}$ increases from $\ang{0.05}$ to $\ang{0.5}$, $\sigma_{D}$ also rises from $\SI{4.6}{\percent}$ to $\SI{12.1}{\percent}$.
Similarly, $\sigma_{D}$ changes from $\SI{4.2}{\percent}$ to $\SI{15.3}{\percent}$ when $\sigma_{\rm \ell}$ varies from $\SI{0.5}{\percent}$ to $\SI{5}{\percent}$.
Conversely, as $\mathcal{R}$ improves from $5000$ to $20000$, $\sigma_{D}$ decreases from $\SI{10.9}{\percent}$ to $\SI{2.8}{\percent}$.
{\cblue 
From Equations \ref{eq:continuum_photon_center} and \ref{eq:diffphase}, it follows that the amplitude of the differential phase depends on $\ell$, which is determined by the analysis of eclipsing light curves.
Approximately, we have
\begin{equation}
    \sigma_{D} \sim \sqrt{\sigma_{\ell}^2 + \sigma_{\phi}^2 + \text{ other terms}}.
\end{equation}
In our fiducial scenario, we have $\sigma_{\ell} \sim \SI{0.6}{\percent}$ and $\sigma_{D} \sim \SI{6}{\percent}$. 
Consequently, the contribution of $\sigma_{\ell}$ to $\sigma_{D}$ is minimal, indicating a lack of strong correlation between $D$ and $\ell$ as illustrated in Figure \ref{fig:dis}. 
This further elucidates why $\sigma_{D}$ exhibits lesser dependence on the quality of the light curve data.}

To further explore the error budget of the measurement of the distance, we also include parameters whose uncertainties have a similar dependence on data quality as the distance's uncertainties in Fig. \ref{fig:std1}.
The relative uncertainty of the semi-major axis ($\sigma_{D}$) has a weaker dependence on $\sigma_{\phi}$ and $\sigma_{\rm \ell}$ but a stronger dependence on $\mathcal{R}$ since the physical size is obtained through the integration of velocity information over time.
Its contribution to $\sigma_{D}$ can be seen in $D \sim a / \Delta \theta$, where $\Delta \theta$ is the angular size of the binary measured by differential phase.
Uncertainties of the inclination angle ($\sigma_{\cos i}$) and the position angle mainly depend on $\sigma_{\phi}$ and $\mathcal{R}$.
They affect the measurement of the distance by changing the projection of the binary orbit on the celestial sphere.
Uncertainties of mass ratio ($q$) increases slightly with larger $\sigma_{\phi}$ and $\sigma_{\rm \ell}$ while decreases significantly with higher $\mathcal{R}$.
They can affect the distances of both stars to the mass center and their radial velocities.
Uncertainties of equivalent widths ($\rm EW_1$ and $\rm EW_2$) are very sensitive to $\sigma_{\rm \ell}$ and $\mathcal{R}$, and they can change the angular position of the photocenter by altering the weights from each star in Eq. \ref{eq:diffphase}.

\subsection{Dependence on model parameters}
\begin{figure}[ht!]
    \includegraphics[width=0.95\textwidth]{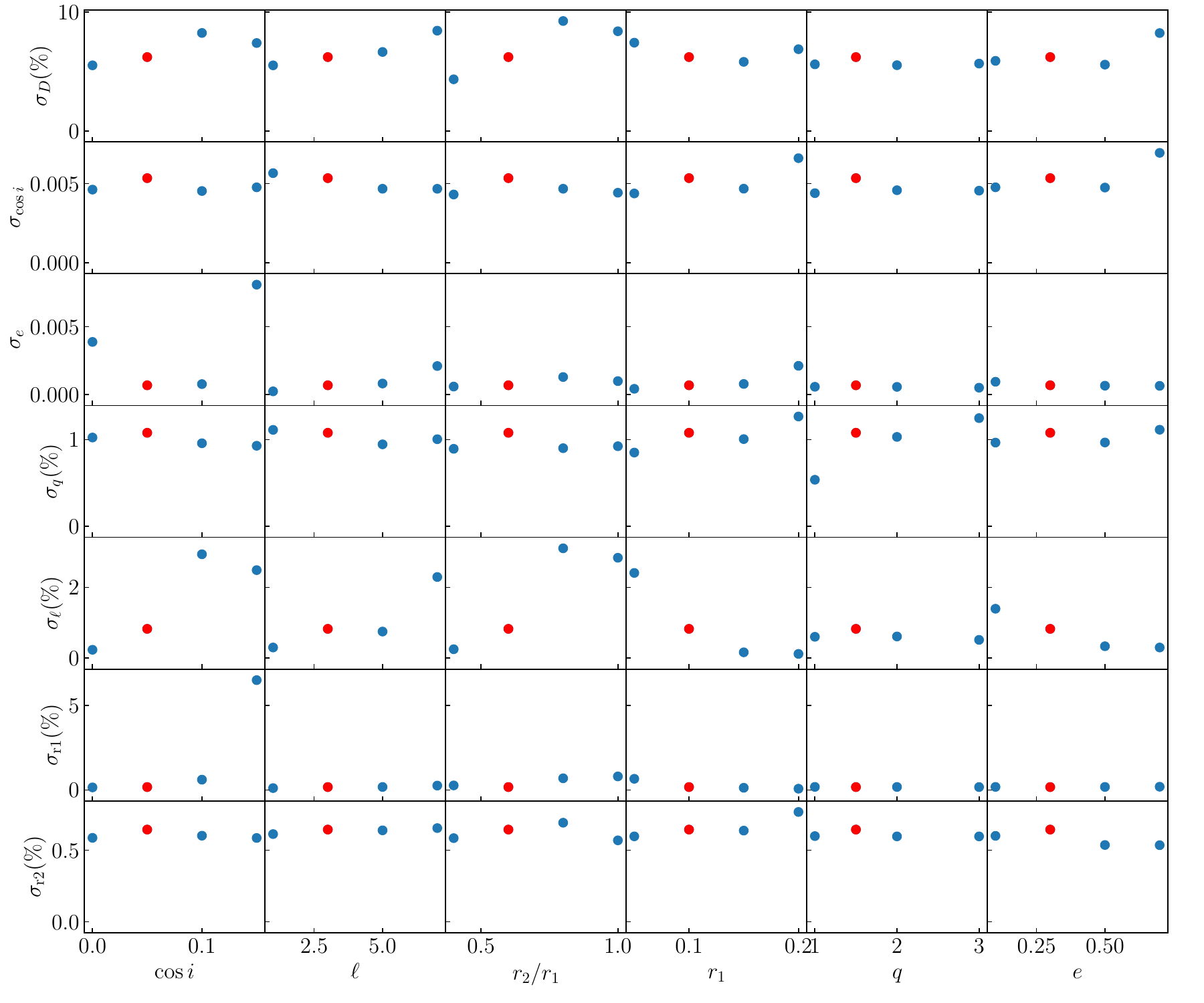}
    \caption{\footnotesize Dependence of measurement uncertainties of some model parameters on their input values.
    The symbols are the same with Figure\,\ref{fig:std1}.
    In each row, we depict the relationship between the uncertainties of a parameter and the input values of itself or other parameters. 
    The uncertainties of distances slightly increase with smaller inclinations, larger luminosity ratios, and larger size ratios, ranging from $\SI{4}{\percent}$ to $\SI{9}{\percent}$.
    The dependence on star size, mass ratio, and orbital eccentricity is very weak.
    }
    \label{fig:std2}
\end{figure}

Now we will explore the impact of model parameters' values on their uncertainties.
Most dimensional parameters, such as $D$, $a$, and $P$, will only adjust the scale of the data curves, and so their effects are equivalent to that of the data quality.
We will focus on those dimensionless parameters which can change the shape of the data curves.
The simulation results are displayed in Fig. \ref{fig:std2}.
Note that when we vary the relative size of the secondary star ($r_2$), we keep the size of the primary ($r_1$) fixed.
But when we change $r_1$, we keep the ratio $r_2 / r_1$ fixed.

As shown by the first row of Fig. \ref{fig:std2}, $\sigma_{D}$ does not vary a lot as we alter the values of some model parameters and only increased slightly with smaller inclination (larger $\cos i$), higher light ratios ($\ell$), larger $r_2 / r_1$ and smaller $r_1$.
Generally, it ranges from $\SI{4.3}{\percent}$ to $\SI{9.3}{\percent}$ for different values of these parameters.

For comparative analysis, we also illustrate in Fig. \ref{fig:std2} the correlation between the uncertainties of other parameters and their respective input values. 
It is observed that the uncertainties of $\cos i$, $q$, $r_1$, and $r_2$ exhibit constancy within a defined range of input values.
Notably, the uncertainty in the $\ell$ showcases a similar dependency on the input values of $\cos i$, $\ell$, $r_2/r_1$, and $r_1$ as observed for the $D$.
A decrease in the inclination angle of the stellar size weakens the eclipsing effect, consequently leading to a reduction in the precision of $\ell$ measurements, which in turn increases the uncertainty of the distance measurement.

\section{Discussions}
\subsection{\cblue Systematic errors}
Our geometric method for determining distances of binaries is independent of empirical relations, eliminating bias from calibrations and sample selections. {\cblue In this paper, we briefly outline the major sources of errors rather than accurately quantify them.
The systematic uncertainties associated with this method primarily originate from the model bias related to the geometry of the components, including factors such as shapes and limb darkening effects when fitting observational data. 
Limb darkening can produce complex yet regular brightness distributions that vary depending on stellar types, necessitating careful parameterization during the analysis of actual data.}
This phenomenon will also influence the stellar angular diameters derived from interferometric measurements. 
Normally, the mean disparity between a uniform disk and a limb-darkening disk's size is about {\cblue $\SI{6}{\percent}$ for $\sim 1000$ stars listed in JMMC Measured Stellar Diameters Catalogue \citep{duvert2016}}.
Given that the stellar angular diameter comprises just $\SI{10}{\percent}$ of the orbital semi-major axis {\cblue for binaries with $\sim \SI{1}{au}$ separations}, the distance measurement error from this phenomenon should be below $\SI{0.6}{\percent}$, significantly lower than the statistical error. 
Nonetheless, as future observational precision advances, this phenomenon will gain prominence, necessitating its consideration.

On the other hand, the intrinsic spectral line profiles of individual stars may also play a crucial role in influencing the values of orbital parameters derived from SA observations. 
Within our model, the individual line profile is intricately shaped by the combined effects of thermal broadening and the rotational characteristics of the star. 
Fig. \ref{fig:dis} illustrates that the thermal broadening velocities, projected rotational velocities, and the orientations of the rotation axes cannot be precisely constrained by the data, suggesting that the distance measurement is not highly reliant on the model employed for line profiles.
However, when stars are sufficiently separated and the orbital velocity falls below the width of an individual star's line profile, the impact of the line shape becomes significantly pronounced.
Additionally, the presence of extended stellar winds in massive stars introduces further complexity, as the inclusion of emission lines from these winds in SA observations may escalate uncertainties in distance measurements \citep{waisberg2017}.

{\cblue To more accurately assess the magnitude of systematic errors introduced by model selection, we propose applying the method to nearby binary star systems with known distances of various types, encompassing a range of characteristics such as colors, shapes, metallicity and other relevant attributes, for interferometric geometric measurements.
Through this comparative analysis, we aim to identify the type of binary star that minimizes systematic errors, thereby optimizing measurements for future extragalactic binary distance determinations.}

\subsection{Spectral energy distribution}
As demonstrated in Equation \ref{eq:diffphase}, the angular displacement of the photocenter is intricately linked to the luminosity ratio existing between the two stars in the binary system. 
Consequently, it is imperative to integrate eclipsing light curves into the dataset for precise determination of the luminosity ratio.
During practical observations, if the stellar size is much smaller compared to the orbital semi-major axis or if the line of sight's inclination angle is more face-on, the weakening of the eclipsing effect can introduce heightened uncertainty in measuring the luminosity ratio. 
In such scenarios, the inclusion of the spectral energy distribution (SED) of the binary system becomes essential. 
Through the fitting and decomposition of these SEDs, the individual temperatures of the stars and their corresponding luminosity ratio can be derived \citep{jadhav2021,thompson2021}.
While absolute flux calibration of the SED holds lesser significance in this approach, a meticulous examination of extinction across different wavelengths remains crucial for obtaining dependable measurements of the luminosity ratio.

\subsection{\cblue Feasibility to determine the LMC distance}
{\cblue In this subsection, we will discuss the feasibility of using existing interferometric equipment, particularly the next-generation beam combiner GRAVITY+ on VLTI \citep{gravityplus2022}, to measure the distances of eclipsing binary systems in the LMC.}
For this purpose, we chose eclipsing binaries of EA type in the LMC from the OGLE Collection of Variable Stars \citep{udalski2015} with I-band magnitudes brighter than $18$ and orbital periods spanning from $100$ to $1000$ days. 
Subsequent cross-referencing with the GAIA database \citep{jimenez-arranz2023} revealed a total of $265$ binaries. 
Of these, $163$ binaries possess K-band magnitudes obtained from the 2MASS survey \citep{skrutskie2006}.

Table \ref{tab:targets_para} presents $17$ binaries in the sample with radial velocities observations \citep{graczyk2018,hoyman2020}.
These systems are composed of two red giants, most of which are G- or early K-type giant stars.
The K-band magnitudes of these systems range from $13$ to $15$, and their orbital parameters are consistent with those of our mock data, making them {\cblue possible candidates} for future {\cblue interferometric} observations.

\begin{figure}
    \centering
    \plotone{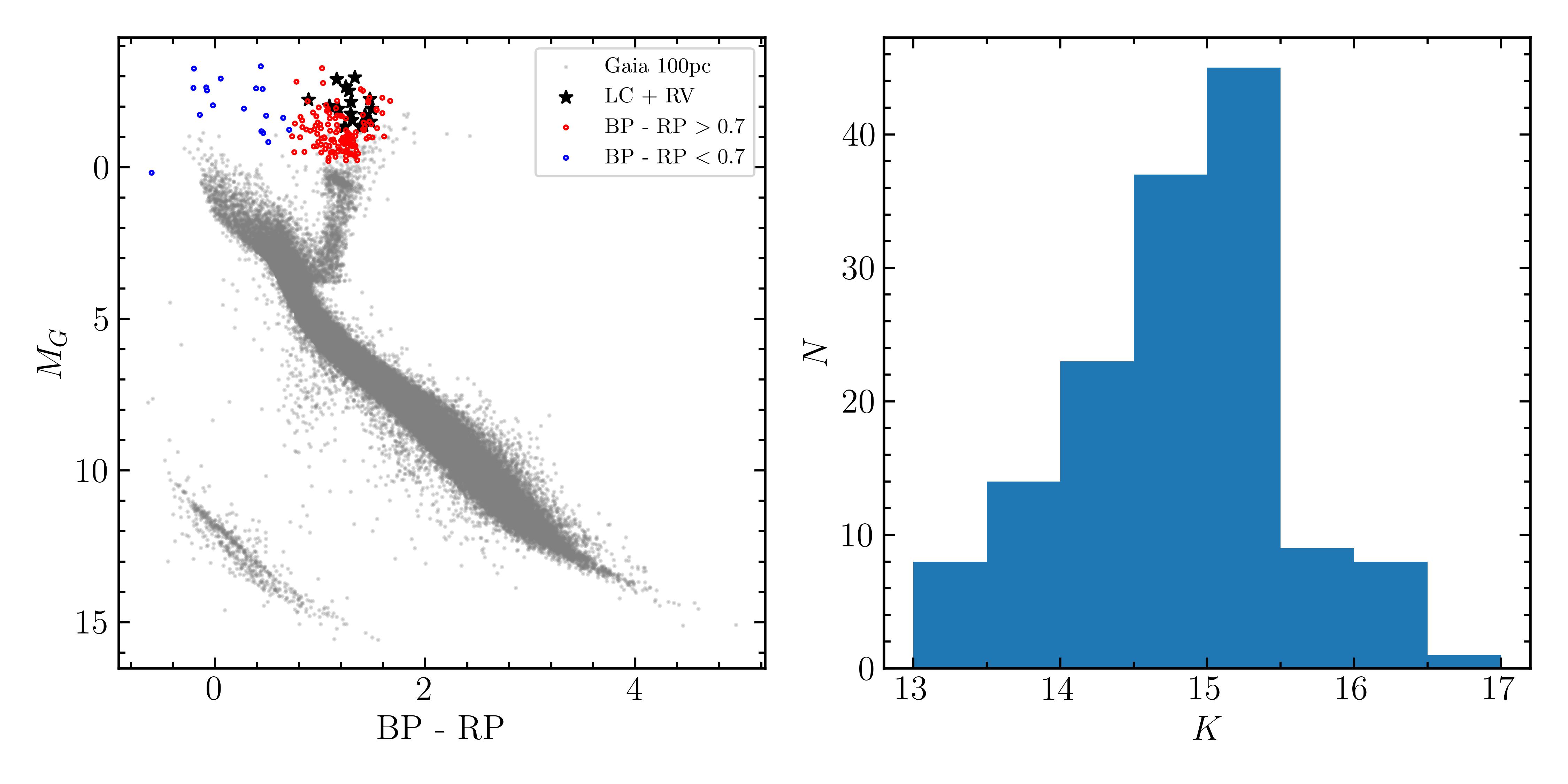}
    \caption{\footnotesize Properties of selected eclipsing binaries in LMC.
    In the left panel, we plot the colors and absolute G band magnitudes of 163 selected targets in LMC.
    The black stars are those with radial velocity curves, composed of two red giants.
    The red circles are the remaining targets with ${\rm BP - RP} > 0.7$, which may have similar properties as those with known orbital parameters.
    The blue circles are the targets with ${\rm BP - RP} < 0.7$, likely composed of O stars or B stars.
    The gray dots are stars from the GAIA database with distances less than $\SI{100}{pc}$ for comparison.
    In the right panel, we show the distribution of K-band magnitude distribution of selected targets with ${\rm BP - RP} > 0.7$. 
    There are $22$ targets with $K < 14$ and $74$ targets with $K < 15$.}
    \label{fig:targets}
\end{figure}

We also show the colors and G band absolute magnitudes of the $163$ selected binaries from LMC in the left panel of Fig. \ref{fig:targets}.
$17$ binaries with known orbital parameters are marked by black stars and located at the red giant branch of the H–R diagram.
For the remaining targets, the red circles represent redder ones with ${\rm BP - RP} > 0.7$.
They may have similar orbital parameters as those 17 targets and be appropriate targets for GRAVITY+ if they are bright enough.
The blue circles are targets with ${\rm BP - RP} < 0.7$.
They are likely composed of O stars or B stars, and we need further spectroscopy data to justify whether they can be used for distance measurements.
In the right panel of Fig. \ref{fig:targets}, we present the distribution of K band magnitude for selected targets with ${\rm BP - RP} > 0.7$.
There are $22$ targets with $K < 14$ and $74$ targets with $K < 15$.

{\cblue The number of photons received by the interferometer per spectral channel for a target can be esitimated by \citep{campins1985}
\begin{equation}
    N_{\rm phot} \approx 1.2 \times 10^{5-0.4(K-13)} \left(\frac{D}{\SI{8.2}{m}}\right)^2 \left(\frac{\Delta t}{\SI{10}{hr}}\right) \left(\frac{\mathcal{R}}{\num{e4}}\right)^{-1} \left(\frac{\eta}{\SI{1}{\percent}}\right),
\end{equation}
where $K$ is the K band magnitude of the target, $D$ is the diameter of the aperture, $\Delta t$ is the duration of the exposure, $\mathcal{R}$ is the spectral resolution and $\eta$ is the overall throughput of the instrument.
Under the best condition, the measurement uncertainty of the phase at each spectral channel will be about  \citep{shao1988}
\begin{equation}
    \sigma_{\phi} = N_{\rm phot}^{-1/2} \si{rad} \approx 0.17 \left(\frac{N_{\rm phot}}{\num{1.2e5}}\right)^{-1/2} \deg.
\end{equation}
Ideally, the differential phase curve of the bightest target in the Table \ref{tab:targets_para} has an amplitude of $1-\SI{2}{\degree}$ and can be obtained using the GRAVITY+ on VLTI with a SNR of $5-10$.
However, we acknowledge that additional noise sources, such as atmospheric pistons and mechanical vibrations, pose significant challenges to observations.
As a first step, we suggest selecting long-period eclipsing binary systems with $K \sim 10$ and $D \sim \SI{5}{kpc}$ within the Milky Way to verify the purely geometric method for distance, as their theoretical SNRs for interferometric phases can exceed $100$. 
Then we will decide whether to observe eclipsing binaries in the LMC using existing interferometers or to wait for future instruments with longer baselines, larger apertures, or higher throughputs, such as unprecedented interferometers combining ELT and VLTI ($\sim 10\,$km baselines) in the next decade.}


\section{Conclusions}
    In this paper, a {\cblue purely geometric} method is developed for determining the distance to extragalactic binaries by conducting a joint analysis of SA, RV, and LC observations. 
    For a typical eclipsing binary in the LMC with a separation of $\SI{2}{au}$ and a period of $\SI{1}{yr}$, observed using optical interferometry with a $\SI{200}{m}$ baseline, the distance uncertainty is approximately $\SI{6}{\%}$ when the precision of interferometric phase reaches $\ang{0.1}$ and the spectral resolution reaches $10000$. 
    The dependence of distance uncertainties on data quality and binary characteristics was systematically analyzed by varying the errors of simulated data and the input values of model parameters. 
    It was observed that the distance measurement precision of individual binary star systems is generally better than $\SI{10}{\percent}$ within a specific range of data quality and input parameters. 
    As a geometric method based on the simplest dynamics, it is independent of empirical calibration and {\cblue the systematics caused by model selections can be tested using nearby binaries with known distances}.
    Geometric distance measurements of nearby galaxies with higher precision can be obtained by measuring multiple binary star systems or monitoring one binary system repeatedly.

\begin{table}
\centering
\footnotesize
\caption{List of {\cblue binary candidates in LMC} with known orbital parameters  \label{tab:targets_para}}
\begin{tabular}{lllcccccccc}
\hline\hline
ID$^a$ & RA & DEC & $K$ & $P$ (day) & $a$ (au) & $e$ & $q$ & $\ell$ $^b$ & $r_1$ & $r_2$ \\
\hline
06575 & 05h04m32.88s & -69d20m51.03s & 13.056 & 189.997 & 1.302 & 0.000 & 1.045 & 1.097 & 0.157 & 0.167 \\
01866 & 04h52m15.28s & -68d19m10.24s & 13.228 & 251.265 & 1.499 & 0.241 & 1.003 & 0.651 & 0.083 & 0.146 \\
33491 & 05h27m00.65s & -67d29m09.87s & 13.586 & 737.991 & 2.947 & 0.325 & 1.005 & 1.161 & 0.010 & 0.010 \\
09660 & 05h11m49.47s & -67d05m45.13s & 13.657 & 167.789 & 1.081 & 0.050 & 1.005 & 2.064 & 0.102 & 0.191 \\
13360 & 05h20m59.46s & -70d07m35.29s & 13.687 & 262.438 & 1.606 & 0.000 & 1.028 & 1.549 & 0.088 & 0.114 \\
10567 & 05h14m01.90s & -68d41m17.91s & 13.740 & 117.959 & 0.879 & 0.000 & 1.047 & 0.516 & 0.130 & 0.194 \\
05430 & 05h01m51.77s & -69d12m48.80s & 13.801 & 505.054 & 2.268 & 0.192 & 1.242 & 1.446 & 0.059 & 0.071 \\
15260 & 05h25m25.55s & -69d33m04.54s & 13.809 & 157.469 & 0.812 & 0.000 & 1.019 & 0.497 & 0.133 & 0.242 \\
03160 & 04h55m51.48s & -69d13m47.92s & 13.993 & 150.157 & 0.847 & 0.000 & 1.006 & 2.640 & 0.093 & 0.205 \\
12875 & 05h19m45.40s & -69d44m38.50s & 14.093 & 152.848 & 0.865 & 0.000 & 1.015 & 3.790 & 0.084 & 0.219 \\
09114 & 05h10m19.56s & -68d58m12.01s & 14.111 & 214.330 & 1.309 & 0.036 & 1.031 & 1.350 & 0.094 & 0.067 \\
18836 & 05h32m53.07s & -68d59m12.23s & 14.322 & 182.531 & 1.122 & 0.107 & 1.027 & 0.511 & 0.066 & 0.128 \\
21873 & 05h39m51.20s & -67d53m00.59s & 14.351 & 144.181 & 0.983 & 0.008 & 1.037 & 1.313 & 0.096 & 0.117 \\
25658 & 06h01m58.78s & -68d30m55.12s & 14.441 & 192.786 & 1.076 & 0.373 & 1.000 & 0.593 & 0.093 & 0.119 \\
12933 & 05h19m53.70s & -69d17m20.38s & 14.471 & 125.380 & 0.710 & 0.000 & 1.001 & 0.334 & 0.113 & 0.239 \\
09678 & 05h11m51.79s & -69d31m01.13s & 14.612 & 114.458 & 0.786 & 0.000 & 1.062 & 2.966 & 0.081 & 0.181 \\
02197 & 04h53m14.68s & -67d33m59.05s & 15.049 & 199.757 & 1.098 & 0.116 & 1.110 & 2.500 & 0.107 & 0.066 \\
\hline
\end{tabular}
\begin{list}{}{}
    \item[Note:~$^a$]{The ID here is an abbreviation for OGLE-LMC-ECL-XXXXX.}
    \item[$^b$]{The $\ell$ here represents the bolometric luminosity ratio calculated from the temperatures and radii of both stars. 
    However, this ratio is close to the luminosity ratio at the observed wavelength since the temperatures of both stars are similar.}
\end{list}
\end{table} 

\vspace{1cm}
Zhanwen Han is acknowledged for useful discussions on binary stars and comments on the first version of the paper. 
YYSS thanks for the support of the fellowship of China National Postdoctoral Program for Innovative Talents through grant BX20230196 and the fellowship from the
China Postdoctoral Science Foundation through grant 2023M732033.
JMW thanks the support of the National Key R\&D Program of China through grant-2016YFA0400701 and -2020YFC2201400 by NSFC-11991050, -11991054, -11833008, -11690024, and by grant No. QYZDJ-4 SSW-SLH007 and No. XDB23010400.

\bibliography{ms}{}
\bibliographystyle{aasjournal}

\end{document}